\documentclass[letterpaper,10pt,conference]{ieeeconf}
\usepackage{graphicx}
\usepackage{color}

\IEEEoverridecommandlockouts                              

\usepackage{cite}
\usepackage{amsfonts,amsmath,amssymb}
\usepackage{enumerate}
\usepackage{comment}

\makeatletter
\let\NAT@parse\undefined
\makeatother
\usepackage{hyperref}%
\hypersetup{colorlinks=true, linkcolor=blue, breaklinks=true, urlcolor=blue, citecolor=blue}

\title{Coalescing Force of Group Pressure:\\Consensus in Nonlinear Opinion Dynamics}

\author{Iryna Zabarianska and Anton V. Proskurnikov
\thanks{I. Zabarianska is with Intelligent Systems Department, Moscow Institute of Physics and Technology.
Anton V. Proskurnikov is with Department of Electronics and Telecommunications at Politecnico di Torino, Turin.}
\thanks{Emails: \texttt{akshiira@yandex.ru,anton.p.1982@ieee.org}}
\thanks{This study was carried out within the 2022K8EZBW ``Higher-order interactions in social dynamics with application to monetary networks'' project—
funded by European Union—Next Generation EU within the PRIN 2022 program
(D.D. 104—02/02/2022 Ministero dell’Universit\'a e della Ricerca). This manuscript
reflects only the authors’ views and opinions and the Ministry cannot be considered
responsible for them.}
}

\def\be{\begin{equation}}
\def\ee{\end{equation}}
\def\ben{\begin{equation*}}
\def\een{\end{equation*}}

\newtheorem{corollary}{Corollary}
\newtheorem{theorem}{Theorem}
\newtheorem{assum}{Assumption}

\newtheorem{definition}{Definition}

\newcommand{\dfb}{\stackrel{\Delta}{=}}
\def\V{\mathcal V}

\def\endofthm{\hfill$\star$}

\usepackage{subcaption}
\begin{document}

\maketitle

\begin{abstract}
This work extends the recent opinion dynamics model from~\cite{cheng2019opinion}, emphasizing the role of group pressure in consensus formation. We generalize the findings to incorporate social influence algorithms with general time-varying, opinion-dependent weights and multidimensional opinions, beyond bounded confidence dynamics. We demonstrate that, with uniformly positive conformity levels, group pressure consistently drives consensus and provide a tighter estimate for the convergence rate. Unlike previous models, the common public opinion in our framework can assume arbitrary forms within the convex hull of current opinions, offering flexibility applicable to real-world scenarios such as opinion polls with random participant selection. This analysis provides deeper insights into how group pressure mechanisms foster consensus under diverse conditions.
\end{abstract}

\section{Introduction}

Models of opinion formation have recently attracted considerable attention from the research community~\cite{XiaWangXuan:2011,friedkin2015problem,proskurnikov2018tutorial,mastroeni2019agent}. While the full range of factors shaping opinions and actions remains under-explored, several simplified models have been proposed to capture the specific effects of social interactions. Since French's foundational work on social power~\cite{1956_french} and Abelson's studies on factors influencing unanimity and disagreement~\cite{abelson1967mathematical}, much focus has been placed on understanding the causes of opinion consensus and divergence.

One factor that positively influences consensus is iterative averaging, where an agent's opinion adjustments tend to keep it within the convex combination spanned by the agent's and their peers' previous opinions.  Currently, the principle of iterative averaging is supported by experimental evidence, ranging from small-group studies~\cite{FriedProsk,FRIEDKINPROBULLO:2021} to analyses of opinions on social media~\cite{Kozitsin2022}.
Reaching consensus through iterative averaging has been studied since the seminal works on rational decision-making published in the 1970s~\cite{Degroot_74,Lehrer:1976}; however, these works assume that the static weighted graph of social influence with certain connectivity properties.

Consensus is easily disrupted by time-varying or opinion-dependent (co-evolving) structures of social influence, as demonstrated by the well-known dynamics of bounded confidence~\cite{Bernardo2024survey,Ceragioli2021,Hegselmann2023survey}. Despite extensive research on averaging algorithms over time-varying graphs (see, e.g.,~\cite{2019_XiaShi,ProCala2023,FrascaFagnaniBook} and references therein), a significant gap remains between necessary and sufficient conditions, with the latter being challenging to test in practice. Another well-understood reason for disagreement is the influence of external signals at each stage of opinion dynamics. These signals can be the initial opinions shaped by the group's previous experience~\cite{friedkin2015problem}, the static opinions of stubborn individuals (``radicals'')~\cite{hegselmann2015opinion}, or the access some agents have to external information~\cite{XiaCao:11}.
In models of iterative averaging with randomized interactions, such constant signals may not only lead to disagreement but also cause sustained oscillations in opinions~\cite{AcemogluComo2013}.

In social psychology, various forms of \emph{group pressure} are well-known factors that promote consensus by driving individuals to align their opinions with the majority~\cite{myers2022social,cialdini1984influence,aronson1972social}. 
The famous experiments by Asch on conformity~\cite{asch1955opinions} demonstrate
that individuals often conform to the majority opinion in a group, even when that opinion was obviously wrong, and tend to prioritize group consensus over personal correctness to avoid standing out or being socially rejected. The effects of group (peer) pressure have also been observed in online social networks~\cite{niu2013,bapna_umyarov2015}. At the same time, there are relatively few dynamical models that reveal the exact mechanisms by which group pressure facilitates consensus.

Among binary opinion models, the most well-known are Granovetter's threshold model~\cite{granovetter1978threshold} and 
Galam's majority rule models~\cite{galam1986majority,galam2008sociophysics,galam2002minority}, which portray the conformity through the majority influence principle: individuals tend to adopt the opinion held by the majority (or at least a certain proportion) of their peer group. Such models are typically analyzed using methods from statistical mechanics.

For continuous opinions, conformity and group pressure are frequently modeled using the Friedkin-Johnsen model~\cite{friedkin2015problem} and its game-theoretic reformulations~\cite{2014_auto_ghaderi}. On one hand, an agent's conformity is characterized by their susceptibility to social influence; in game reformulations, this parameter in the utility function represents a penalty for disagreement with others. This interpretation was further developed in~\cite{semonsen2019pressure,ferraioli2019pressure}, demonstrating that if the penalty weights are time-varying and gradually increasing, the best-response consensus dynamics eventually converge to consensus. Informally, disagreement between agents' opinions leads to increasing discomfort, which pressures them to abandon individual prejudices and seek unanimity with the group. On the other hand, group pressure can be modeled as individuals' reluctance to express their genuine opinions, avoiding overt disagreement with the majority. This concept is captured in the EPO (Expressed-Private Opinion) models~\cite{ye2019EPO}, an extension of the Friedkin-Johnsen model, where each agent has private and expressed opinions, with only the latter displayed to peers.

A completely different approach to social pressure was proposed in~\cite{cheng2019opinion,cheng2020consensus}. The models developed there demonstrate how group pressure enables consensus to reemerge in bounded confidence opinion dynamics, which would otherwise result in opinion clustering.
The model from~\cite{cheng2019opinion} assumes that agents have access to a common public opinion, which they incorporate at each stage of their opinion update. The influence weight assigned to the public opinion reflects an agent's willingness to align with the group and characterizes their level of conformity. If these weights are positive, consensus is achieved in finite time. The extension proposed in~\cite{cheng2020consensus} incorporates the mechanism of expressed and private opinions, similar to that in~\cite{ye2019EPO}, thereby merging two different approaches to modeling group pressure.

This work extends the results of~\cite{cheng2019opinion} in multiple directions. 

\textbf{First}, the ``underlying'' opinion dynamics need not necessarily be classical Hegselmann-Krause bounded confidence; we allow arbitrary averaging algorithm with time-varying and opinion-dependent influence weights and vector-valued opinions. We show that, as soon as all agents have non-zero conformity levels, the coalescing force of social pressure necessary drives the group to consensus. The peculiarity of bounded confidence dynamics is that consensus is achieved in finite time, whereas, in general, consensus is asymptotic.

\textbf{Second}, the structure of the common "public" opinion can be arbitrary, as long as it lies within the convex hull of the current opinions, and it is not required to be the mean of all opinions, unlike~\cite{cheng2019opinion}. This flexibility is crucial, as the "averaged" opinion of a real social group is typically derived from opinion polls with a random selection of participants, and the data from such polls can be processed further, e.g., by removing outlier opinions or assigning them low weights.

\textbf{Third}, we significantly improve the convergence time estimate for the model. We prove that the time required to reach $\varepsilon$-consensus depends only on $\log\varepsilon$ and the minimal conformity level of the agents, and is independent of the group size $n$. 
For the bounded confidence influence mechanism, this also provides an estimate for the termination time.
Notably, the conformity coefficients do not need to be static.

The remainder of the paper is organized as follows. Section~\ref{sec.model} introduces the model and related concepts.
Section~\ref{sec.main} presents our main result, giving conditions for consensus caused by public pressure.
Section~\ref{sec.simul} illustrates this result by numerical simulations. Section~\ref{sec.concl} concludes the paper.


\section{Opinion Dynamics with Group Pressure}\label{sec.model}

The model introduced in this section integrates general dynamics of iterative averaging with group pressure as in~\cite{cheng2019opinion}.

\subsubsection{\bf Agents and their opinions} 
Consider a group of agents denoted by $\V \dfb\{1, \ldots, n\}$. Each agent $i$ is characterized by a vector-valued\footnote{A vector opinion can represent an individual's attitude toward a complex object or problem, capturing multiple facets that cannot be conveyed by a single scalar value, such as in resource allocation problems among multiple entities~\cite{friedkin2019mathematical} or attitudes toward several interrelated or independent topics, as in belief system modeling~\cite{Parsegov2015CDC}. Multidimensional opinions naturally arise in various sociodynamic models~\cite{friedkin2015problem,PedrazaBalenzuela2021}.} \emph{opinion}, being a  
function $\xi^{i}(t) = (\xi^{i}_{1}(t), \ldots, \xi^{i}_{d}(t))\in \mathbb{R}^d$ of the discrete time $t = 0,1,\ldots$.
We use symbol $\Xi(t)$ to denote the state of the system, i.e., the family of opinions $\Xi(t)=(\xi^i(t))_{i\in\mathcal{V}}\in\mathbb{R}^{nd}$.

\subsubsection{\bf Underlying Opinion Dynamics} Following the approach in~\cite{cheng2019opinion}, we consider some opinion dynamics based on iterative averaging, which, unlike~\cite{cheng2019opinion}, is not limited to a bounded confidence model. This system, which can be nonlinear and non-stationary, may lead to opinion disagreement and even sustained oscillations in them, as seen, e.g., in our recent model proposed in~\cite{zaba2024setbased}. 
Henceforth, this auxiliary dynamical system, described by the general equations
\begin{equation}\label{eq.underlying}
\begin{gathered}
\xi^{i}(t+1)=\sum\nolimits_{j\in\V}w_{ij}(t,\Xi(t))\xi^j(t),\\
w_{ij}(t,\Xi)\geq 0,\;\;\sum\nolimits_{j\in\V}w_{ij}(t,\Xi)=1,
\end{gathered}
\end{equation}
will be referred to as the \emph{underlying} opinion dynamics. Typically, matrices $W(t,\Xi(t))$ are sparse as each agent interacts only to few peers. We do not require any connectivity properties for the digraphs associated with these matrices.

\subsubsection{\bf Public opinion} Similar to~\cite{cheng2019opinion,cheng2020consensus,cheng2022social}, we assume that agents are aware of a \emph{public} opinion, which represents an averaged view of the group as a whole. In previous works~\cite{cheng2019opinion,cheng2020consensus,cheng2022social}, the public opinion is taken as the simple mean of all opinions; however, obtaining this value is not always feasible. In online social network analysis, opinions are typically mined through sentiment analysis~\cite{liu2012sentiment}, making it challenging to capture the opinions of billions of users in a short timeframe.
Opinions of ``offline'' social groups are typically analyzed through opinion polls, with participants selected randomly and representing only a small percentage of the group. Furthermore, the collected data can be preprocessed, for example, to reject outlier or marginal opinions or assign them lower weights.

Hence, we assume that the public opinion, provided to the agents by some "oracle," is a function (generally time-varying and nonlinear)
$\xi^{pub}(t)=\xi^{pub}(t,\Xi(t))$, which performs some form of opinion averaging in the following sense.
\begin{assum}\label{asm.pub}
For all time instants $t=0,1,$ and all opinion vectors $\Xi=(\xi^1;\ldots;\xi^n)\in\mathbb{R}^{nd}$, one has
\[
\xi^{pub}(t,\Xi)\in {\rm conv}\{\xi^1,\ldots,\xi^n\},
\]
where symbol ${\rm conv}$ denotes the convex hull\footnote{The \emph{convex hull} of vectors $\xi^1,\dots, \xi^n\in\mathbb{R}^d$ is the polytope
\begin{gather}
conv(\xi^1,\dots, \xi^n) \dfb \left\{\sum_{i\in\V}\alpha_i\xi^i : \sum_{i\in\V}\alpha_i = 1,\quad \alpha_i \ge 0
\right\}\subset\mathbb{R}^d.
\end{gather}}.\endofthm
\end{assum}

\subsubsection{\bf Opinion Formation Model} 

The opinion updates proceed in two steps: \emph{local} opinion averaging followed by \emph{alignment} with the public opinion.

At \textbf{Step 1}, each agent aggregates the opinions of others as defined by the underlying opinion formation model~\eqref{eq.underlying}, computing the new opinion
\begin{equation}\label{eq.avg-local}
\eta^i(t) \dfb \sum\nolimits_{j\in\V}w_{ij}(t,\Xi(t))\xi^j(t),\quad i\in\V.
\end{equation}

At \textbf{Step 2}, agents incorporate the public opinion, representing the group as a whole, thereby aligning with the collective opinion of the majority.
The resulting opinion of each agent $i\in\V$ is given by
\begin{equation}\label{eq.pub_press}
\xi^i(t+1)=p_{i}(t)\xi^{pub}(t,\Xi(t)) + (1 - p_{i}(t))\eta^i(t),
\end{equation}
Here, $p_i(t) \in [0,1]$ is the coefficient representing the agent's level of conformity, indicating their willingness to align with the group's opinion.
If all agents are non-conformists in the sense that $p_1=\ldots=p_n\equiv 0$, the model~\eqref{eq.pub_press} reduces to the underlying opinion dynamics~\eqref{eq.underlying}. We will consider the opposite situation and examine the convergence to consensus in the case where all coefficients $p_i$ are
positive, that is, all agents tend to align with the group's opinion.



\begin{definition}
 We call the system~\eqref{eq.avg-local},\eqref{eq.pub_press} the opinion dynamics with group pressure (ODGP) associated with the dynamics of iterative averaging~\eqref{eq.underlying} and the public pressure function $\xi^{pub}=\xi^{pub}(t,\Xi)$.
\end{definition}

\subsection*{A special case from~\cite{cheng2019opinion}}

The model introduced in~\cite{cheng2019opinion} (see also Model I in~\cite{cheng2022social}) is a special case of the ODGP, where
\begin{equation}\label{eq.mean-value}
\xi^{pub}(t,\Xi)=\xi^{pub}(\Xi)\dfb\frac{\xi^1+\ldots+\xi^n}{n}
\end{equation}
is the mean value of all opinions, and the underlying averaging dynamics is the Hegselmann-Krause (HK) model~\cite{2002_hk} with heterogenous confidence radii $\varepsilon_1,\ldots,\varepsilon_n>0$, where\footnote{We use symbol $|N|$ to denote the cardinality of a finite set $N$ (along with the usual absolute value of a number). Unless otherwise stated, symbol $\|\cdot\|$ denotes the Euclidean norm.}
\begin{equation}\label{eq.W_GHK}
\begin{gathered}
w_{ij}(t,\Xi)=w_{ij}(\Xi)\dfb 
\begin{cases}
  \frac{1}{|\mathcal{N}_{i}(\Xi)|},  & \mbox{if } j \in \mathcal{N}_{i}(\Xi)\\
  0, & \mbox{otherwise}
\end{cases}\quad\text{and}\\
\mathcal{N}_i(\Xi)\dfb\{j\in\V:\|\xi^j-\xi^i\|\leq\varepsilon_i\}
\end{gathered}
\end{equation}
It is well-known that the HK model generally leads to opinion clustering. To our knowledge, the existence of oscillatory solutions in this system remains an open problem~\cite{Bernardo2024survey}.

\section{Consensus Caused by Group Pressure}\label{sec.main}

Our main result, presented in this section, guarantees convergence of all opinions in the ODGP to consensus and extends the main result from~\cite{cheng2019opinion} in several directions.
We start with the definition.
\begin{definition}
The solution to the ODGP system asymptotically converges to consensus if 
a common limit exists 
\[
\lim_{t\to\infty}\xi^1(t)=\ldots=\lim_{t\to\infty}\xi^n(t)\in\mathbb{R}^d.
\]
\end{definition}

The following theorem is the criterion of consensus in the ODGP model. For brevity, denote the minimal level of conformity in the group at time $t$ by
\[
p_*(t)\dfb\min_{i\in\V}p_i(t)\in [0,1].
\]
We will also use the maximal distance between the opinions, or the ``diameter''\footnote{Using the norm's convexity, it is easy to check that $R(\Xi)$ coincides with the diameter of the convex hull $M ={\rm conv}(\xi^1,\dots, \xi^n)$, spanned by the opinions, that is, ${\rm diam}\,M=\max_{x,y\in M}\|x-y\|$.} of the opinion matrix
\[
R(\Xi)\dfb\max_{i,j\in\V}\|\xi^i - \xi^j\|.
\]
For brevity, denote
\[
R_0\dfb R(\Xi(0))=\max_{i,j\in\V}\|\xi^i(0) - \xi^j(0)\|.
\]

\begin{theorem}\label{thm.press_converg}
Assume that the function $\xi^{pub}$ obeys Assumption~\ref{asm.pub}. Then, the following statements are valid:

(i) The ODGP asymptotically converges to consensus for every initial condition if
the following series diverges
\begin{equation}\label{eq.series}
\sum\nolimits_{t=0}^{\infty}p_*(t)=+\infty.
\end{equation}

(ii) If, additionally, the levels of conformity remain uniformly positive $p_*(t)\geq p_0>0$ for all $t\geq 0$, the convergence to consensus is exponential, and
$R(\Xi(t))\leq R_0(1-p_0)^t$.

(iii) If $p_*(t_0)=1$ at some $t_0$, the ODGP terminates at time $t_0+1$, and consensus is established
\[
\xi^1(t)=\ldots=\xi^n(t)\equiv\xi^{pub}(t_0)\;\;\forall t\geq t_0+1.\quad\quad\star
\]
\end{theorem}
\vskip0.5cm
It should be noticed that, depending on the underlying averaging dynamics~\eqref{eq.underlying}, finite-time consensus can be guaranteed even if $p_*(t)<1$ for all $t$. As noticed in~\cite{cheng2019opinion}, this holds, e.g., for the HK dynamics~\eqref{eq.W_GHK} and public opinion~\eqref{eq.mean-value}.

\begin{corollary}\label{cor.HK}
Consider the ODGP with the underlying HK dynamics~\eqref{eq.W_GHK} and let $\varepsilon \dfb \min_i \varepsilon_i$ be the minimal confidence radius. 
Assume also that $\xi^{pub}$ is the mean value~\eqref{eq.mean-value}.
If the series~\eqref{eq.series} diverges, then consensus is reached after a finite number of iterations. Specifically, there exists an instant $T_0$ (depending on $R_0, \varepsilon$, and the sequence $p_*(t)$) such that
\[
\xi^1(t)=\ldots=\xi^{n}(t)\quad\forall t\geq T_0.
\]
If $p_*(t)\geq p_0>0$, the time instant $T_0$ can be estimated as
\begin{equation}\label{eq.t-eps2}
T_0\leq \begin{cases}
1,\quad &\varepsilon\geq R_0\\
1+\lceil\frac{\log\varepsilon-\log R_0}{\log(1-p_0)}\rceil,&\varepsilon<R_0.
\end{cases}
\end{equation}
\end{corollary}

Before proceeding with the proof of Theorem~\ref{thm.press_converg} and Corollary~\ref{cor.HK}, we find it important to discuss their key features.

\subsection{Discussion}

\textbf{First}, it is important to highlight the key difference from the alternative approach to social pressure modeling~\cite{semonsen2019pressure,ferraioli2019pressure}. In those works, consensus is achieved through \emph{gradually increasing} pressure, which, in the case of synchronous updates~\cite{semonsen2019pressure}, drives the opinion dynamics toward standard DeGroot dynamics. The condition~\eqref{eq.series} permits the minimal conformity level to asymptotically vanish $p_*(t)\xrightarrow[t\to\infty]{}0$, meaning the effect of social pressure can potentially \emph{diminish} as the agents' opinions converge to unanimity, providing
that this diminution is ``not too fast''. This may seem counter-intuitive because, when all $p_i$ values approach $0$, the dynamics~\eqref{eq.pub_press} are close to the underlying system~\eqref{eq.underlying}, which, unlike the DeGroot model in~\cite{semonsen2019pressure}, does not guarantee consensus and, in principle, can have oscillatory solutions as, e.g., the modified HK system from~\cite{zaba2024setbased}.

\textbf{Second}, whereas the proof of Theorem~\ref{thm.press_converg} is simple, this result, to the best of our knowledge, is not covered by any existing consensus criterion. The original choice of $\xi^{pub}$ as the ``mean-field'' term~\eqref{eq.mean-value} in~\cite{cheng2019opinion}, along with the assumption that $p_i>0$ are constant, essentially, allows the ODGP to be viewed as an averaging algorithm over a complete weighted graph with uniformly positive, time-varying weights independent of the matrix $w_{ij}$. The conditions of Theorem~\ref{thm.press_converg}, however, allow the graphs associated with $W(t) = (w_{ij}(t, \Xi(t)))$ to be arbitrary and permit \( p_i(t) \to 0 \) for all \( i \) as \( t \to \infty \), making it impossible to validate standard consensus conditions such as repeated quasi-strong connectivity, arc-balance, or cut-balance~\cite{2019_XiaShi,proskurnikov2020recurrent,FrascaFagnaniBook,MatvPro:2013}.
A similar consensus condition to~\eqref{eq.series} appears in~\cite{DeMarzo2003}, but the latter work addresses a distinct linear model and relies heavily on eigenvector decomposition for analysis. Finally, a special case of Theorem~\ref{thm.press_converg}, dealing with a special case where one of the agents is stubborn and $\xi^{pub}$
stands for its public opinion, appears in~\cite{MaiAbed2014}.

\textbf{Third}, comparing the consensus time estimate in Corollary~\ref{cor.HK} with~\cite[Theorem~1]{cheng2019opinion}, two key differences emerge: (a) our result applies to multidimensional opinions, and (b) the convergence time is independent of $n$. Indeed, numerical experiments in the next section confirm that the time required to reach consensus does not increase as the group size grows.

\subsection{Proofs of Theorem~\ref{thm.press_converg} and Corollary~\ref{cor.HK}}

\textbf{Step 1.} Notice first that, as in all iterative averaging algorithms, the convex hulls spanned by the opinions
\[
M(t)\dfb{\rm conv}\{\xi^1(t),\ldots,\xi^n(t)\}
\]
are nested $M(t+1)\subseteq M(t)$. This is immediate from Assumption~\ref{asm.pub} and the equations~\eqref{eq.avg-local},~\eqref{eq.pub_press} and the definition of the convex hull, because $\xi^i(t+1)\in M(t)$ for all $i\in\V$.

In particular, if the opinions are in consensus at some time $t_*$, i.e., $\xi^i(t_*)=\xi^*$ for all $i\in\V$ and some $\xi^*\in\mathbb{R}^d$, then the ODPG terminates at time $t_*$, because $M(t)=\{\xi^*\}$, and
\[
\xi^{pub}(t)=\xi^1(t)=\ldots=\xi^n(t)=\xi^*\quad\forall t\geq t_*.
\]
From this fact, statement (iii) in Theorem~1 is immediate. Indeed, if $p_*(t_0)=1$, then $1\geq p_i(t_0)\geq p_*(t_0)=1$ for all $i$, and hence consensus of the opinions is established at time $t_*=t_0+1$ with $\xi^*=\xi^{pub}(t_0)$.

\textbf{Step 2.} Given a trajectory $\Xi(t)=(\xi^1(t);\ldots;\xi^n(t))$ of the ODGP, let
$R(t)=R(\Xi(t))$.
From Step 1, it is straightforward that $R(t+1)\leq R(t)$.
We will prove a stronger inequality
\begin{equation}\label{eq.aux}
R(t+1)\leq (1-p_*(t))R(t).
\end{equation}
Denote for brevity $\xi^{pub}(t)=\xi^{pub}(t,\Xi(t))$.
Choosing a pairs of agents $i,j\in\V$, assume, without loss of generality, that $p_i(t) \ge p_j(t)$.
Then, subtracting the equations~\eqref{eq.pub_press} for the agents $i$ and $j$, one obtains
\[
\begin{aligned}
\xi^i(t+1) - \xi^j(t+1)=&(p_i(t) - p_j(t))\left(\xi^{pub}(t) - \eta^{j}(t)\right)+\\
+&(1-p_i(t))\left(\eta^{i}(t) - \eta^{j}(t)\right).
\end{aligned}
\]
In view of~\eqref{eq.avg-local} and Assumption~\ref{asm.pub}, we have
\[\xi^{pub}(t),\eta^i(t),\eta^j(t)\in{\rm conv}\{\xi^1(t),\ldots,\xi^n(t)\}.
\]
Since $R(t)$ is the diameter of the latter convex hull, one has
\[
\begin{aligned}
\|\xi^i(t+1) &- \xi^j(t+1)\|\leq (p_i(t) - p_j(t))\left\|\xi^{pub}(t) - \eta^{j}(t)\right\|+\\
&+(1-p_i(t))\|\eta^{i}(t) - \eta^{j}(t)\|\leq\\
&\leq (p_i(t)-p_j(t))R(t)+(1-p_i(t))R(t)\leq\\
&\leq(1-p_j(t))R(t)\leq (1-p_*(t))R(t).
\end{aligned}
\]
The latter inequality holds for every pair of agents $i,j$, which entails~\eqref{eq.aux}. The statement (ii) in Theorem~\ref{thm.press_converg} is now straightforward, since $1-p_*(t)\leq 1-p_0$. 

\textbf{Step 3.} 
To prove statement (i), it suffices, due to Step~1, to consider the case where $p_*(t)<1$ for all $t$.
It is well-known\footnote{See, for instance,~\cite[Chapter~15, Theorem 15.5]{rudin1987real}} that the series diverges~\eqref{eq.series} if and only if
\[
\prod\nolimits_{t=0}^{\infty}(1-p_*(t))=0,
\]
entailing that $R(t)\xrightarrow[t\to\infty]{}0$. Since $M(0)\supseteq M(1)\supseteq\ldots \supseteq M(t)\supseteq\ldots$ is a sequence of non-empty compact sets,  
their intersection is non-empty~\cite[Chapter~2, Theorem~2.6]{rudin1987real} and contains at least one point $\xi^*\in \bigcap_{t\geq 0}M(t)$. Obviously,
\[
0\leq\|\xi^i(t)-\xi^*\|\leq R(t)\xrightarrow[t\to\infty]{}0\quad\forall i,
\]
which proves statement (i).

To prove statement (ii) in Theorem~\ref{thm.press_converg}, it suffices to show that $p_*(t)\geq p_0>0$ entails that $R(t)\leq (1-p_0)^tR_0$, hence,
$\log R(t)\leq \log R(0)+t\log(1-p_0)\log R(0)$. Hence, if $\varepsilon\leq R(0)$, then $\epsilon$-consensus is achieved immediately at time $t=0$.
Otherwise, it is guaranteed when $t\geq\mathcal{T}(\varepsilon,R(0),p_0)$, since $t\log(1-p_0)\leq \log\varepsilon-\log R(0)$ in view of~\eqref{eq.t-eps} and $\log(1-p_0)<0$. This finishes the proof of Theorem~\ref{thm.press_converg}.

\textbf{Step 4.} To prove Corollary~\ref{cor.HK}, note that for every initial condition and every $\varepsilon>0$, there exists a time step $T_{\varepsilon}$ at which 
\emph{$\varepsilon$-consensus} is achieved, i.e., $R(t) \leq \varepsilon$.
Choosing $\varepsilon = \min_i \varepsilon_i$, we have that at time $T_{\varepsilon}$ all agents become neighbors, i.e., $N_i(T_{\varepsilon}) = \V$, as indicated by~\eqref{eq.W_GHK}. Then, by~\eqref{eq.avg-local} and the assumption~\eqref{eq.mean-value}, it follows that $\eta^i(T_{\varepsilon}) = \xi^{pub}(T_{\varepsilon})$ for all $i \in \V$. Hence, consensus is reached at $T_0 \dfb T_{\varepsilon} + 1$, since
\[
\xi^i(T_0) = \xi^{pub}(T_{\varepsilon})\quad\forall i.
\]
This also implies, as demonstrated in Step~1, that the ODPG terminates at time $T_0$. If, additionally, $p_*(t)\geq p_0>0$, then $\log R(t)\leq \log R(0)+t\log(1-p_0)\log R(0)$, which, as can be easily seen, entails that the $\varepsilon$-consensus ($R(t)\leq\varepsilon$) is established no later than at time 
\begin{equation}\label{eq.t-eps}
\mathcal{T}(\epsilon,R_0,p_0)\dfb
\begin{cases}
0,\quad &\varepsilon\geq R_0\\
\lceil\frac{\log\varepsilon-\log R_0}{\log(1-p_0)}\rceil,&\varepsilon<R_0,
\end{cases}
\end{equation}
and the ODPG terminates no later than at time $T_0=1+\mathcal{T}(\epsilon,R_0,p_0)$, which finishes the proof of Corollary~\ref{cor.HK}.

\section{Numerical Simulations}\label{sec.simul}

In this section, we illustrate the behavior of ODGP through several experiments. In all tests, the underlying averaging dynamics is the two-dimensional ($d=2$) Hegselmann-Krause model, corresponding to the influence matrix~\eqref{eq.W_GHK}, with homogeneous confidence radii $\varepsilon_i = \varepsilon > 0$ and levels of conformity $p_i=p>0$. The initial conditions are randomly sampled from the square $[-1,1]\times [-1,1]$.

\subsection*{Different types of the public opinion}

Our first experiment demonstrates that consensus is indeed achieved for various public opinions satisfying Assumption~\ref{asm.pub}, not just for the mean of all opinions~\eqref{eq.mean-value}. In this experiment, we randomly sample a subset $\V_{p}(t) \subseteq \V$ of $k$ participants at each step $t$, $\xi^{pub}(t)$ being their mean value:
\[
\xi^{pub}(t,\Xi)=\frac{1}{k}\sum\nolimits_{j\in V_{p}(t)}\xi^j.
\]
When $k=1$, this ``public'' opinion represents that of a single randomly chosen individual, making it completely non-representative. In the opposite extreme, with $k=n$, the sample mean matches~\eqref{eq.mean-value}.

\begin{figure}[htb]
     \centering     
     \begin{subfigure}[b]{0.49\columnwidth}
         \centering
         \includegraphics[width=\textwidth]{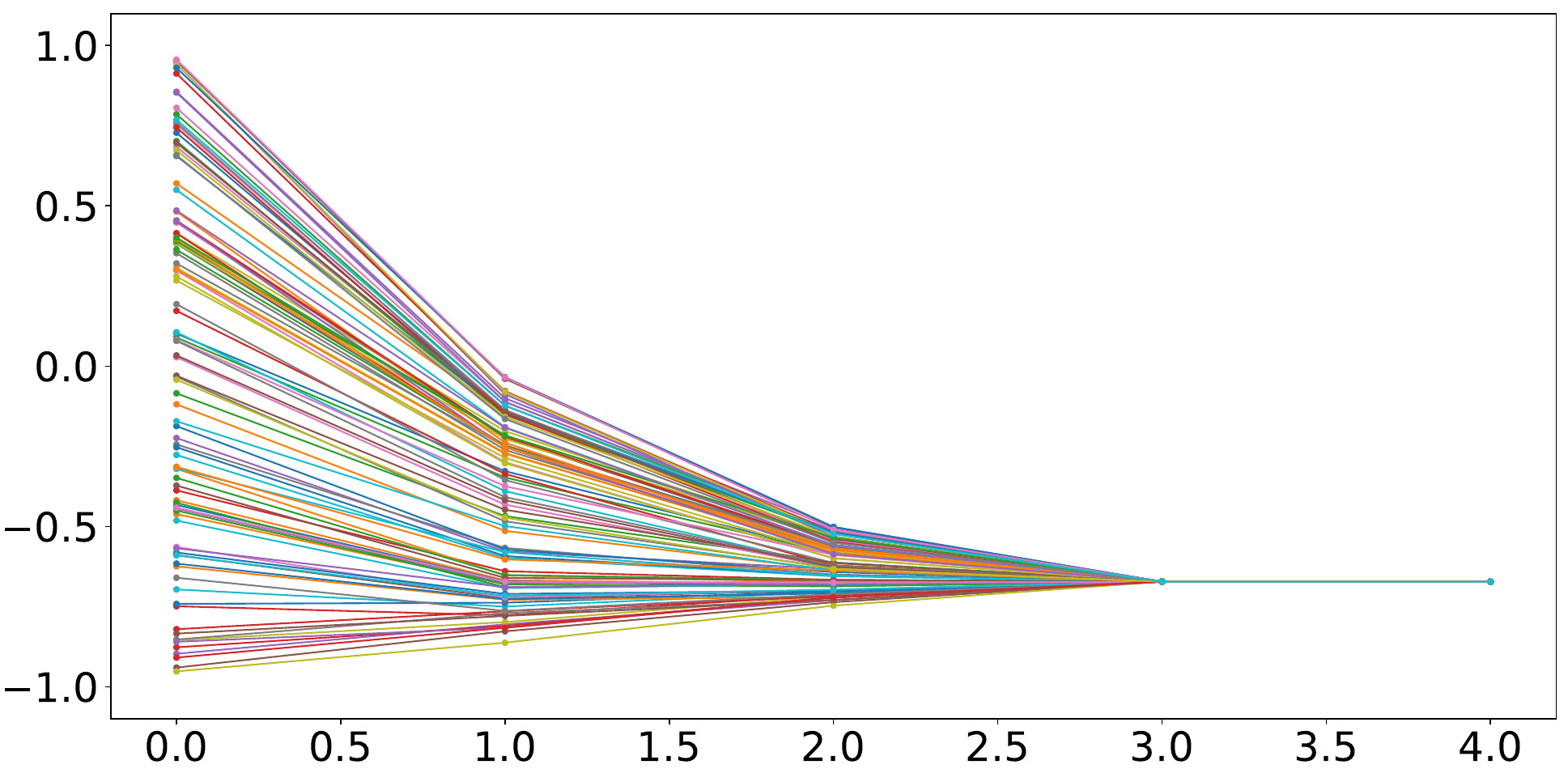}
         \caption{k=1, first coordinate}
    \end{subfigure}
    \hfill
     \begin{subfigure}[b]{0.49\columnwidth}
         \centering
         \includegraphics[width=\textwidth]{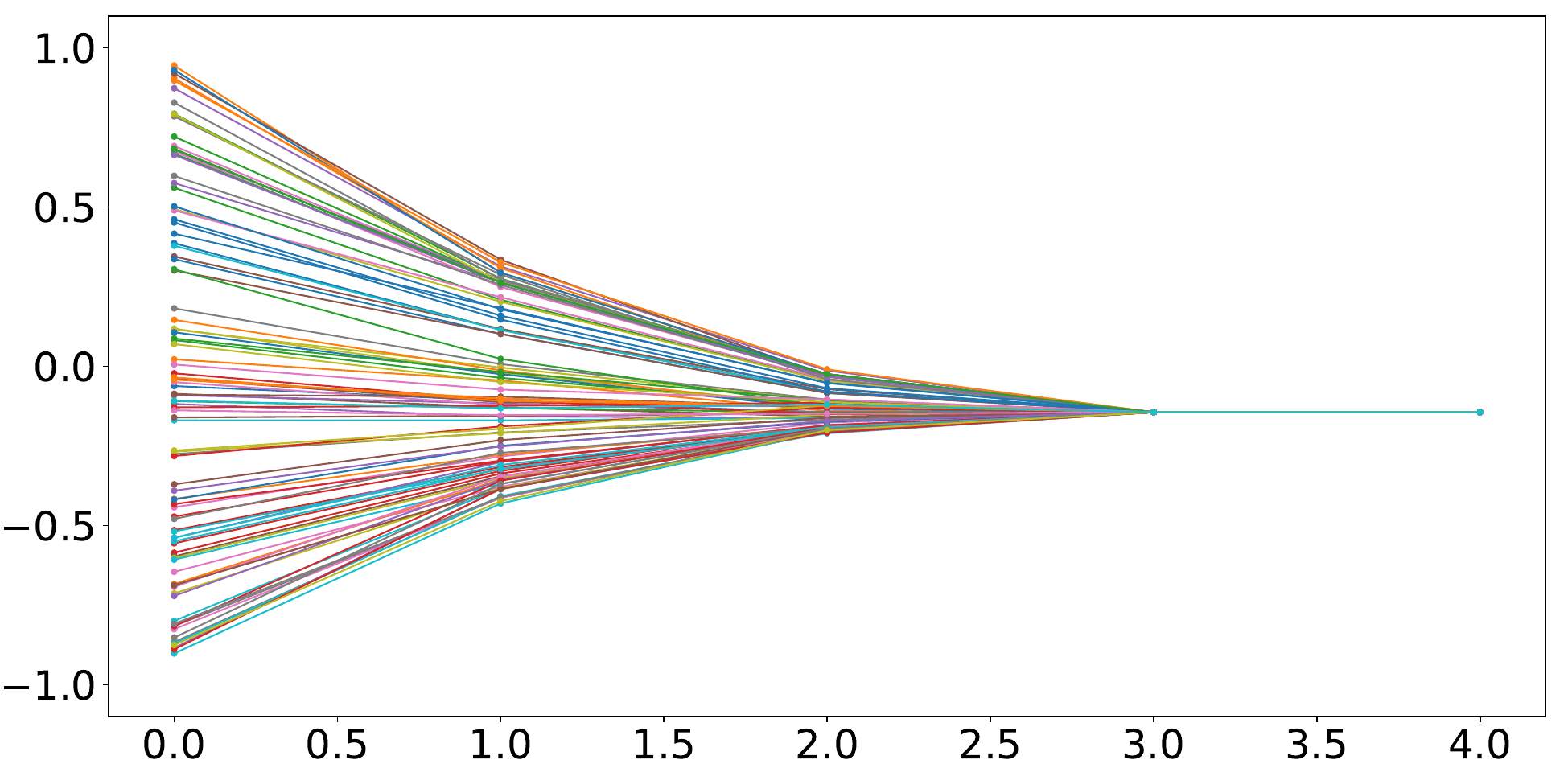}
         \caption{$k=1$, second coordinate}
     \end{subfigure}\\
     \begin{subfigure}[b]{0.49\columnwidth}
         \centering
         \includegraphics[width=\textwidth]{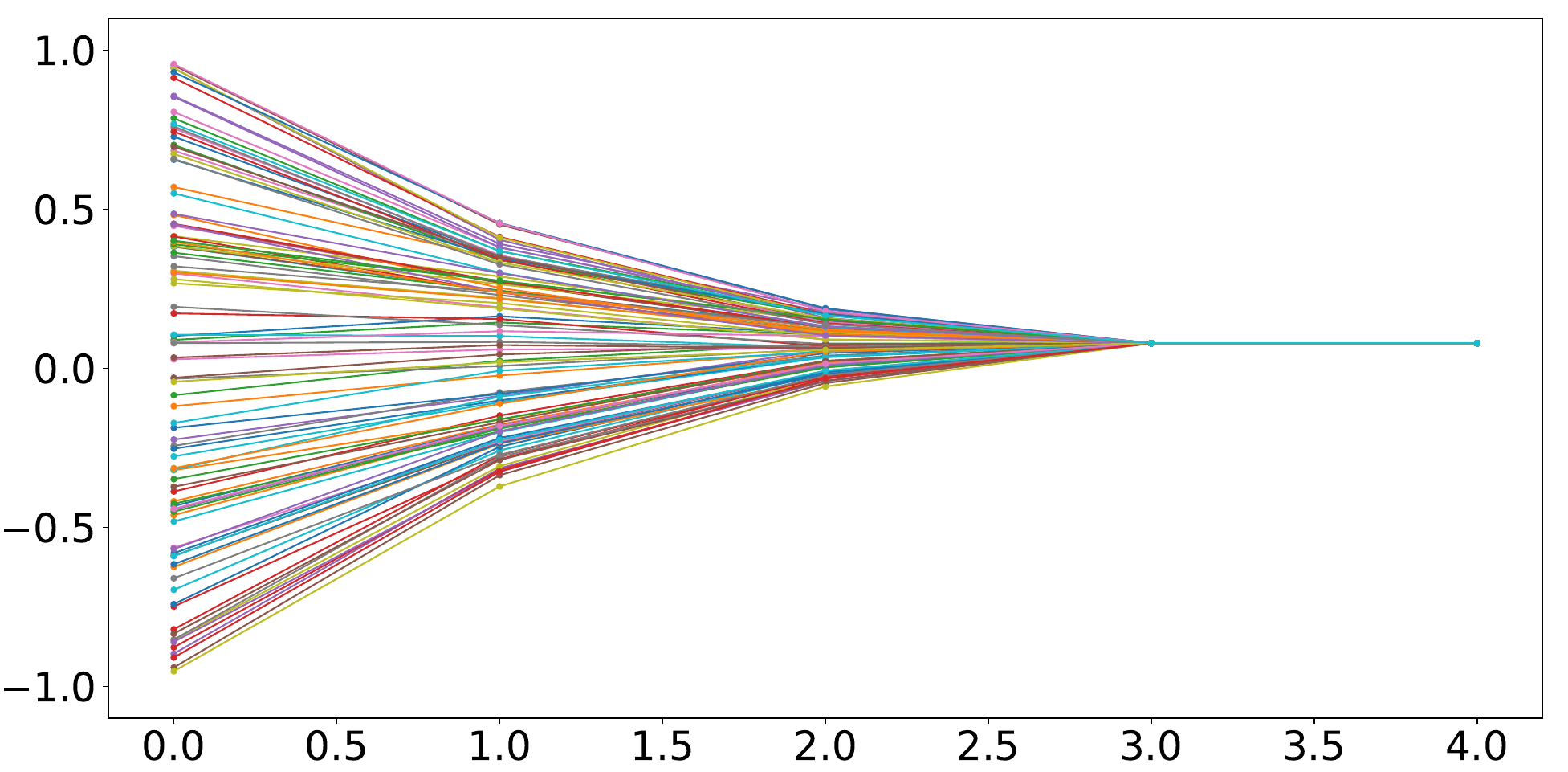}
          \caption{$k=20$, first coordinate}
    \end{subfigure}
    \hfill
     \begin{subfigure}[b]{0.49\columnwidth}
         \centering
         \includegraphics[width=\textwidth]{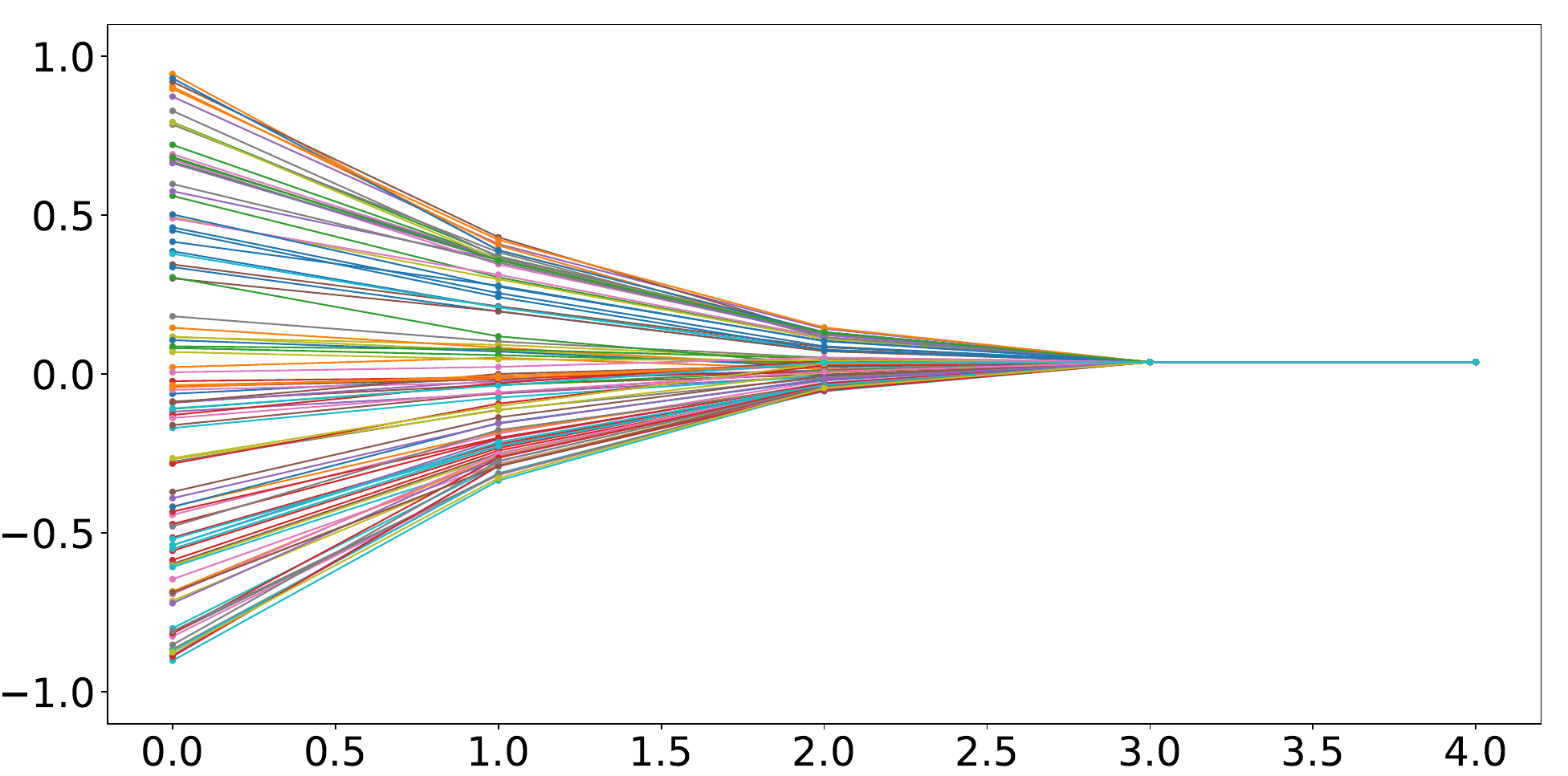}
          \caption{$k=20$, second coordinate}
     \end{subfigure}\\
     \begin{subfigure}[b]{0.49\columnwidth}
         \centering
         \includegraphics[width=\textwidth]{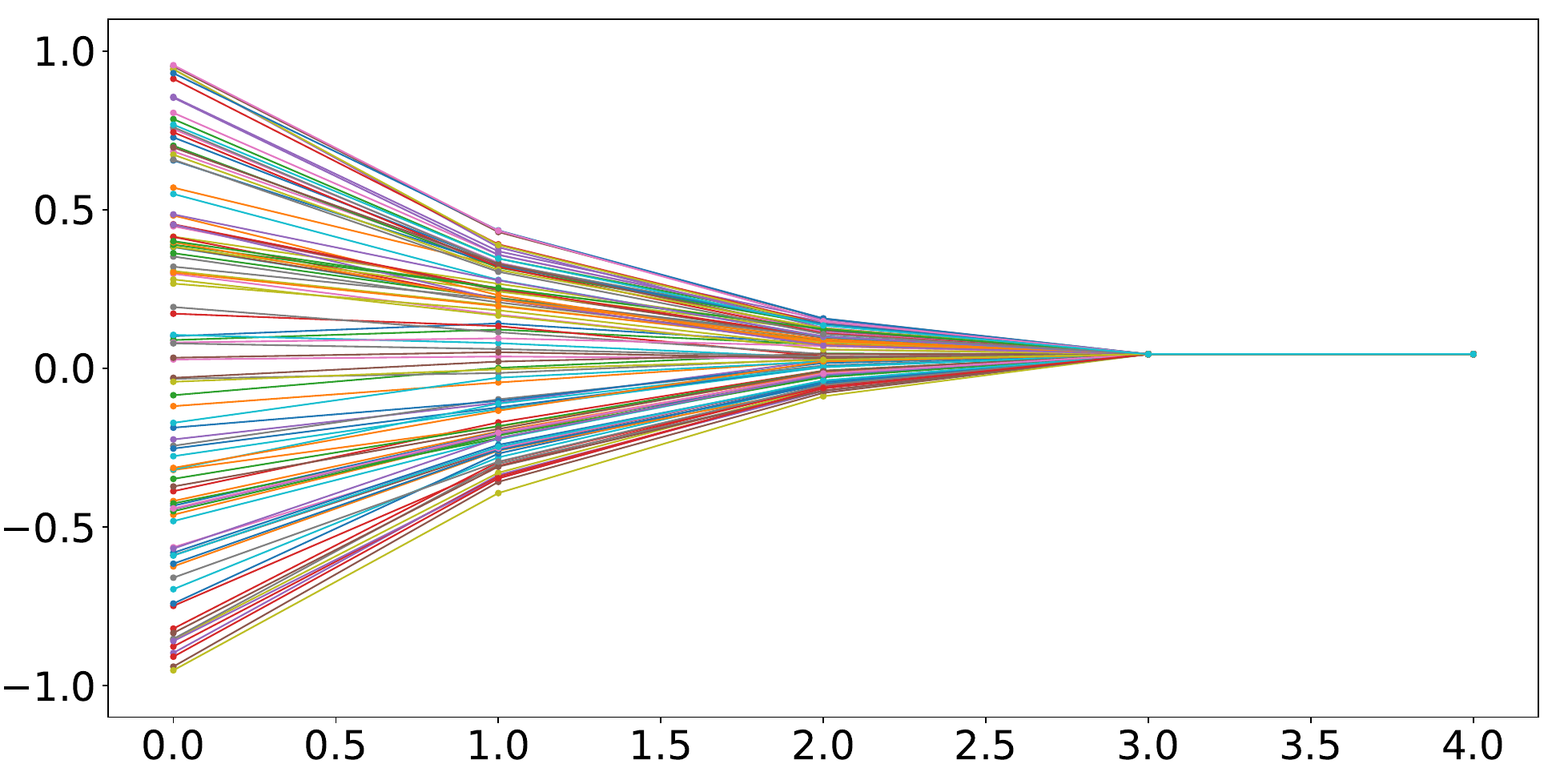}
          \caption{$k=100$, first coordinate}
    \end{subfigure}
    \hfill
     \begin{subfigure}[b]{0.49\columnwidth}
         \centering
         \includegraphics[width=\textwidth]{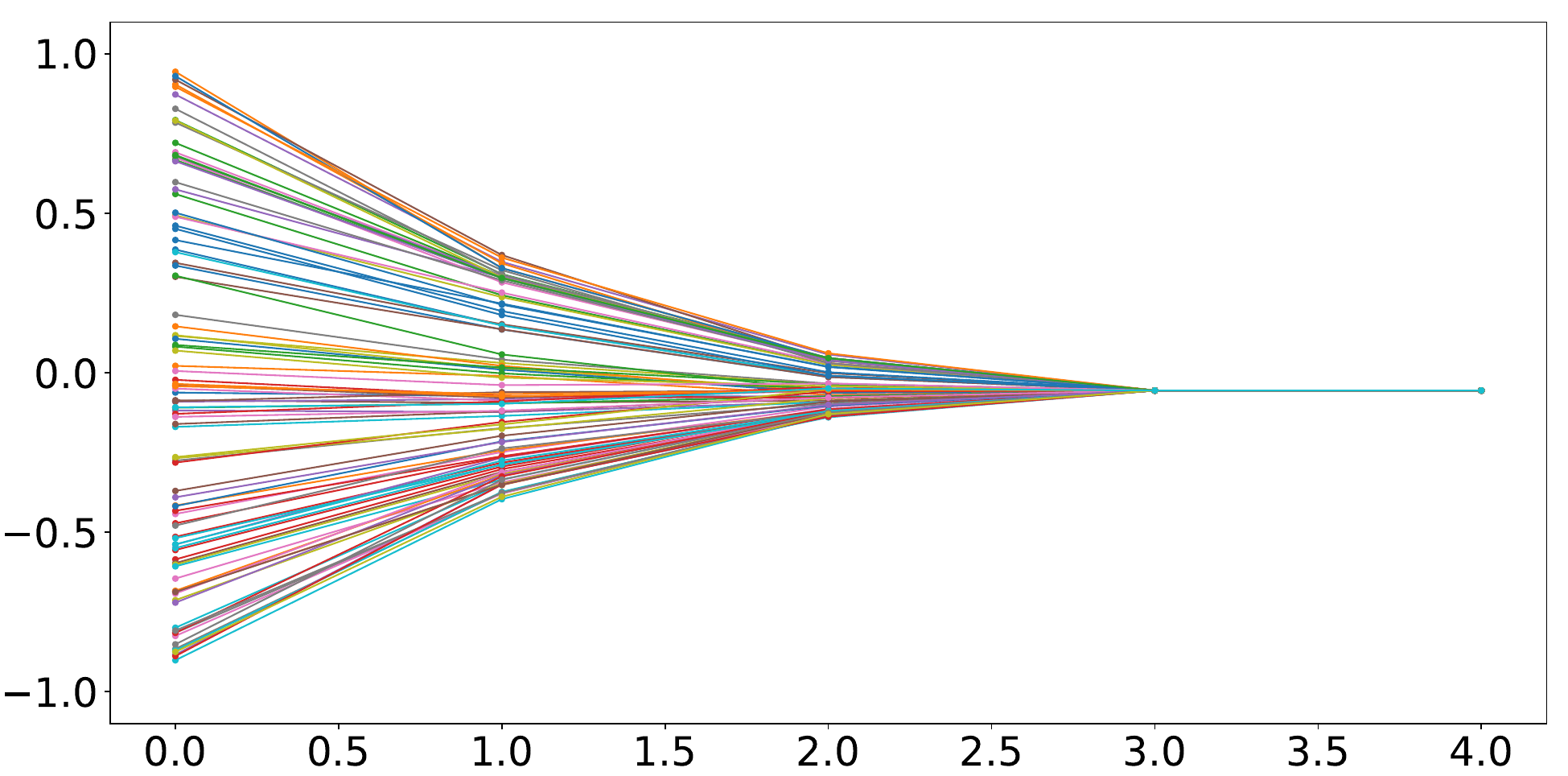}
          \caption{$k=100$, second coordinate}
     \end{subfigure}
     \caption{ODGP with Different $\xi^{pub}(t)$. 
     \textbf{Left part}: dynamics of $\xi_1^i(t)$ (vertical axis) vs. time (horizontal axis); \textbf{Right part}: dynamics of $\xi_2^i(t)$ for $i \in \V$}
     \label{fig.pub-diff}
\end{figure}

We performed three simulations with $k=1, 20, 100$ using the same initial conditions to compare opinion trajectories. For easier comparison, we plot the two coordinates of the opinions separately. In all simulations, we have $n=100,p=0.5,\varepsilon=0.5$.
The results are shown in Fig.~\ref{fig.pub-diff}.
It can be noticed that the different structure of $\xi^{pub}$ does not influence the convergence time,
however, the consensus opinion for $k=1$, as can be expected,
substantially differs from those for $k=20$ and $k=100$ (the latter being quite close).

\subsection*{The convergence time vs. the network size}
\begin{figure*}[htb]
     \centering     
     \begin{subfigure}[b]{0.4\textwidth}
         \centering
         \includegraphics[width=\textwidth]{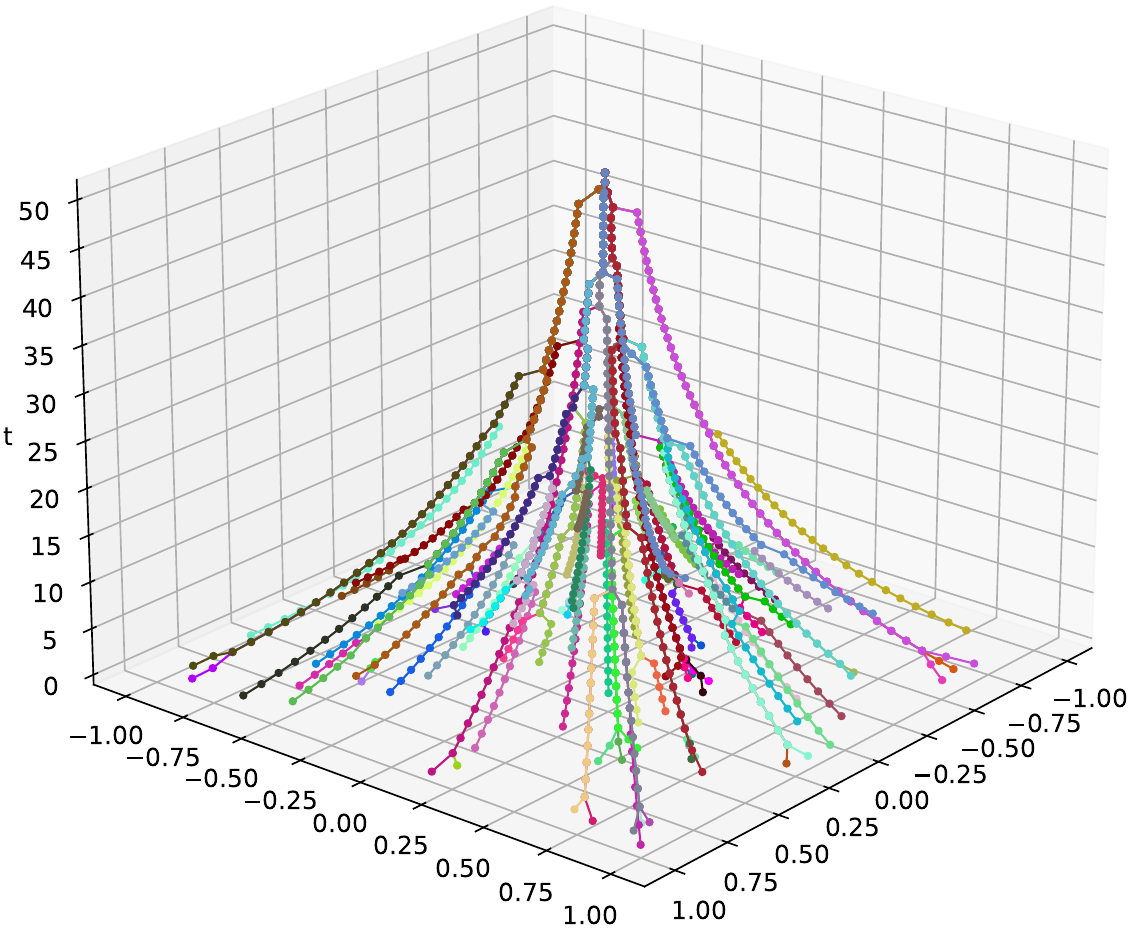}
         \caption{$n=100$, $\varepsilon=0.1$}
    \end{subfigure}
    \hfill
     \begin{subfigure}[b]{0.4\textwidth}
         \centering
         \includegraphics[width=\textwidth]{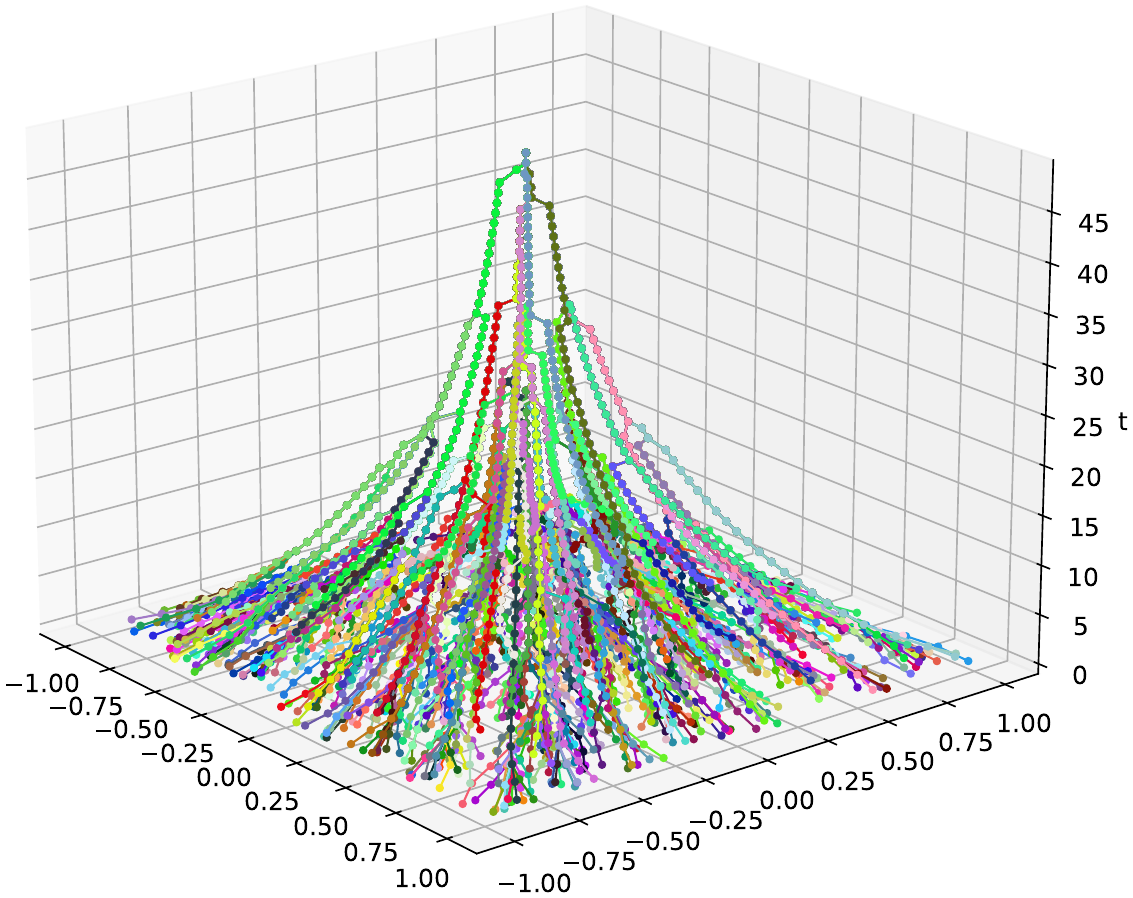}
         \caption{$n=1000$, $\varepsilon=0.1$}
     \end{subfigure}\\
     \begin{subfigure}[b]{0.4\textwidth}
         \centering
         \includegraphics[width=\textwidth]{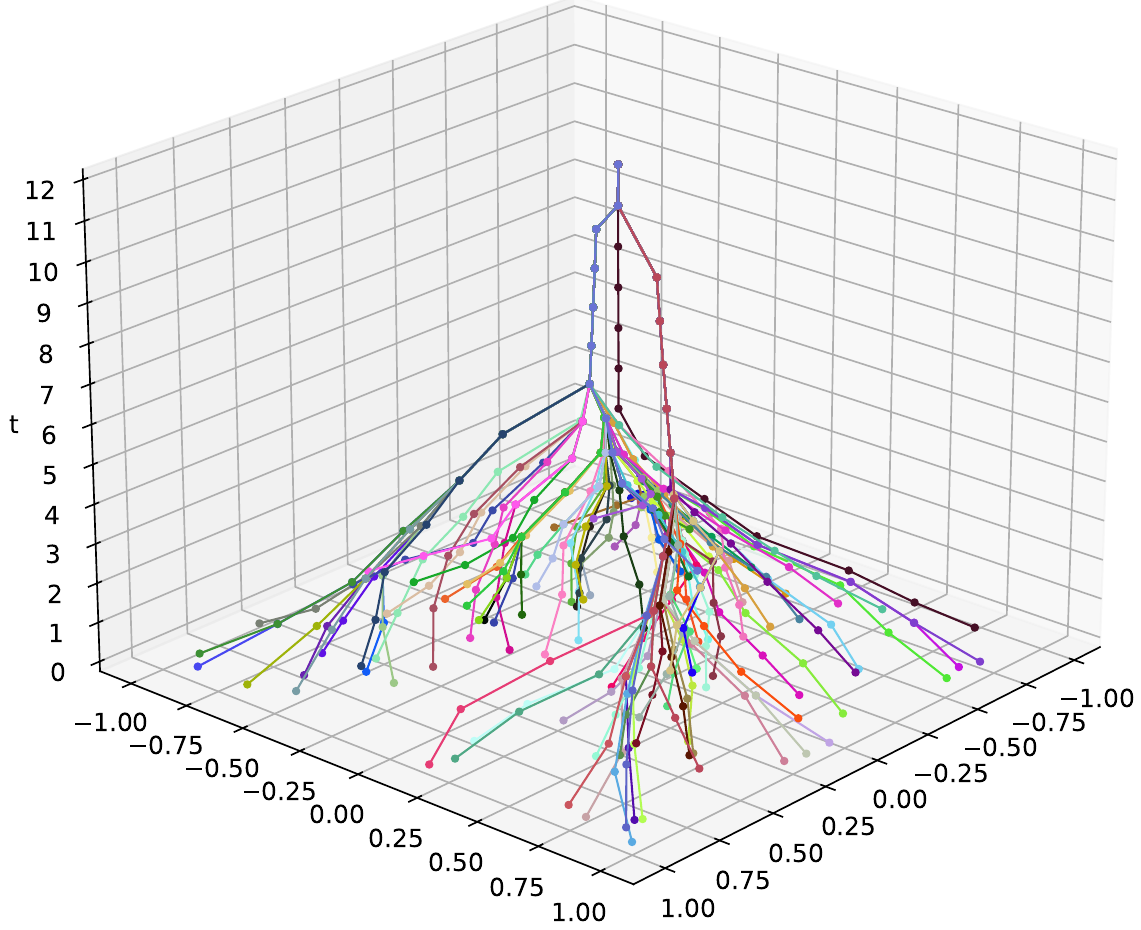}
         \caption{$n=100$, $\varepsilon=0.5$}
    \end{subfigure}
    \hfill
     \begin{subfigure}[b]{0.4\textwidth}
         \centering
         \includegraphics[width=\textwidth]{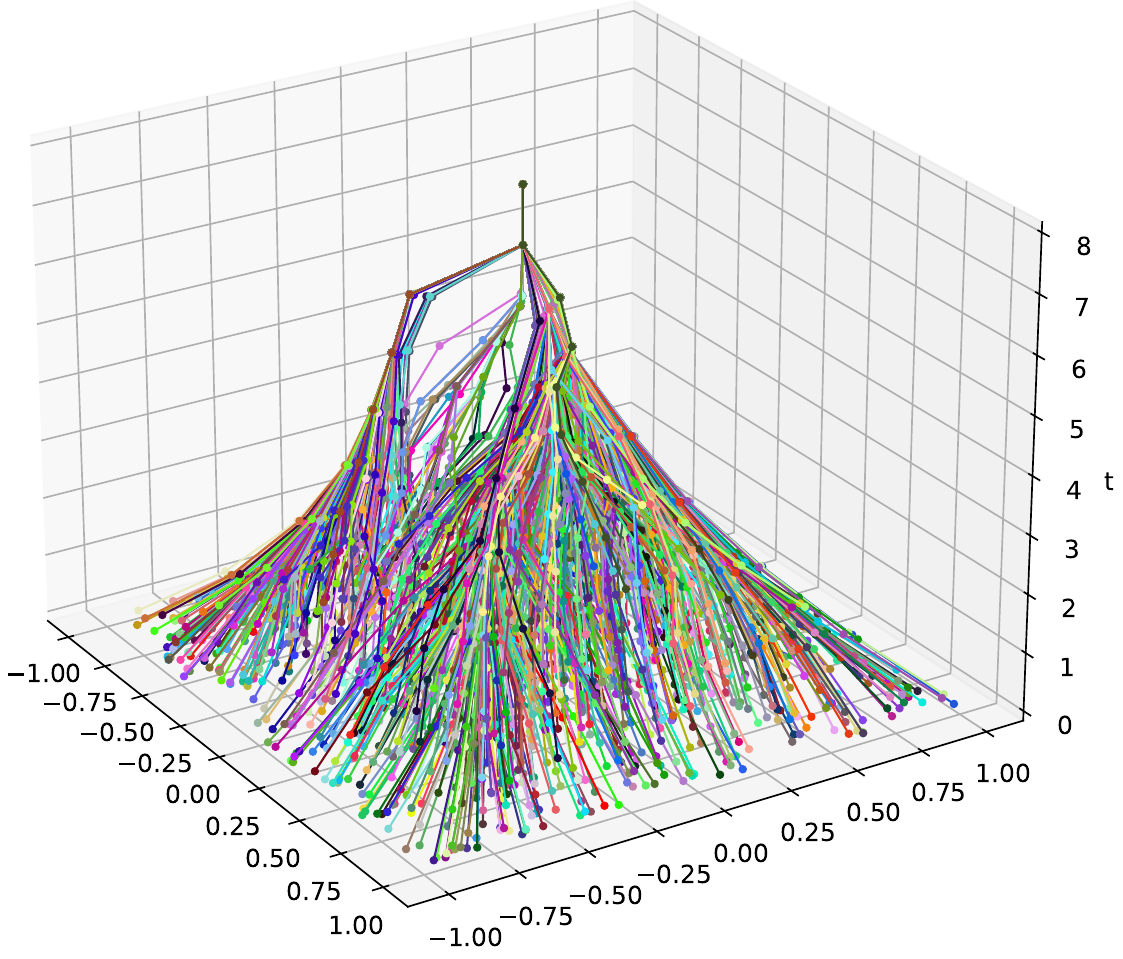}
         \caption{$n=1000$, $\varepsilon=0.5$}
     \end{subfigure}
     \caption{ODGP with different $n$ and confidence radii $\varepsilon$, conformity level is $p=0.05$}
     \label{fig.n}
\end{figure*}

According to Corollary~\ref{cor.HK}, the upper bound for consensus time is independent of $n$, regardless of how small $p_0 = p > 0$ is. 
Our second experiment demonstrates that the number of agents $n$ does not significantly impact the \emph{actual} convergence time, whereas the confidence radius $\varepsilon$ does. 

We started 4 simulations: two for $n=100$ randomly chosen initial opinions (same in the two experiments) with $\varepsilon=0.1$ and $\varepsilon=0.5$, and $n=1000$ randomly chosen opinions (also same in the two experiments), with the same radii.
In all experiments, $p=0.05$, and we use the mean value~\eqref{eq.mean-value} for the public opinion. The opinions are shown in Fig.~\ref{fig.n}
(where the vertical axis is for time).

The initial diameters of the convex hull are similar, with $R_0 \approx 2.635$ for $n=100$ and $R_0 \approx 2.723$ for $n=1000$. The experiments indicate that, for the same $\varepsilon$, the \emph{actual} consensus times for $n=100$ and $n=1000$ are very close (50 steps for $n=100$ versus 48 steps for $n=1000$ when $\varepsilon=0.1$; 11 steps for $n=100$ versus 8 steps for $n=1000$ when $\varepsilon=0.5$). Thus, the actual consensus time does not increase with $n$; in fact, convergence for $n=1000$ appears even faster, likely due to differences in the initial conditions.

It is interesting to compare the actual consensus time with predicted one. For $n=100$, the predicted consensus time is $65$ steps ($\varepsilon=0.1$) and $34$ (for $\varepsilon=0.5$). For $n=1000$, the predicted values are $66$ and $35$.

\subsection*{Convergence time and the conformity level $p$}

Finally, we compare convergence time for different levels of pressure $p$. The predicted consensus time~\eqref{eq.t-eps2} is inversely proportional to $\log(1-p)$, decreasing as $p$ increases. 

\begin{figure*}[htb]
     \centering     
     \begin{subfigure}[b]{0.33\textwidth}
         \centering
         \includegraphics[width=\textwidth]{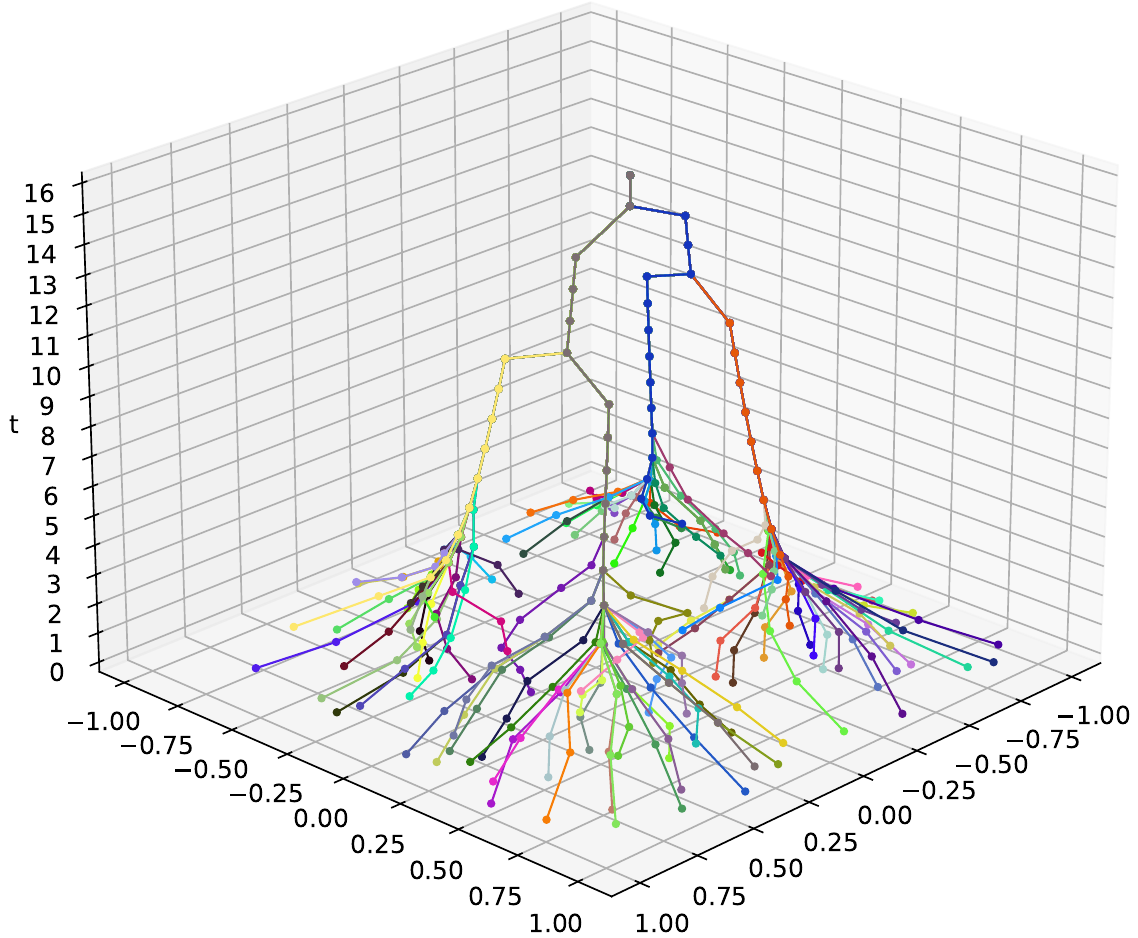}
         \caption{$p=0.05$}
    \end{subfigure}
    \hfill
     \begin{subfigure}[b]{0.33\textwidth}
         \centering
         \includegraphics[width=\textwidth]{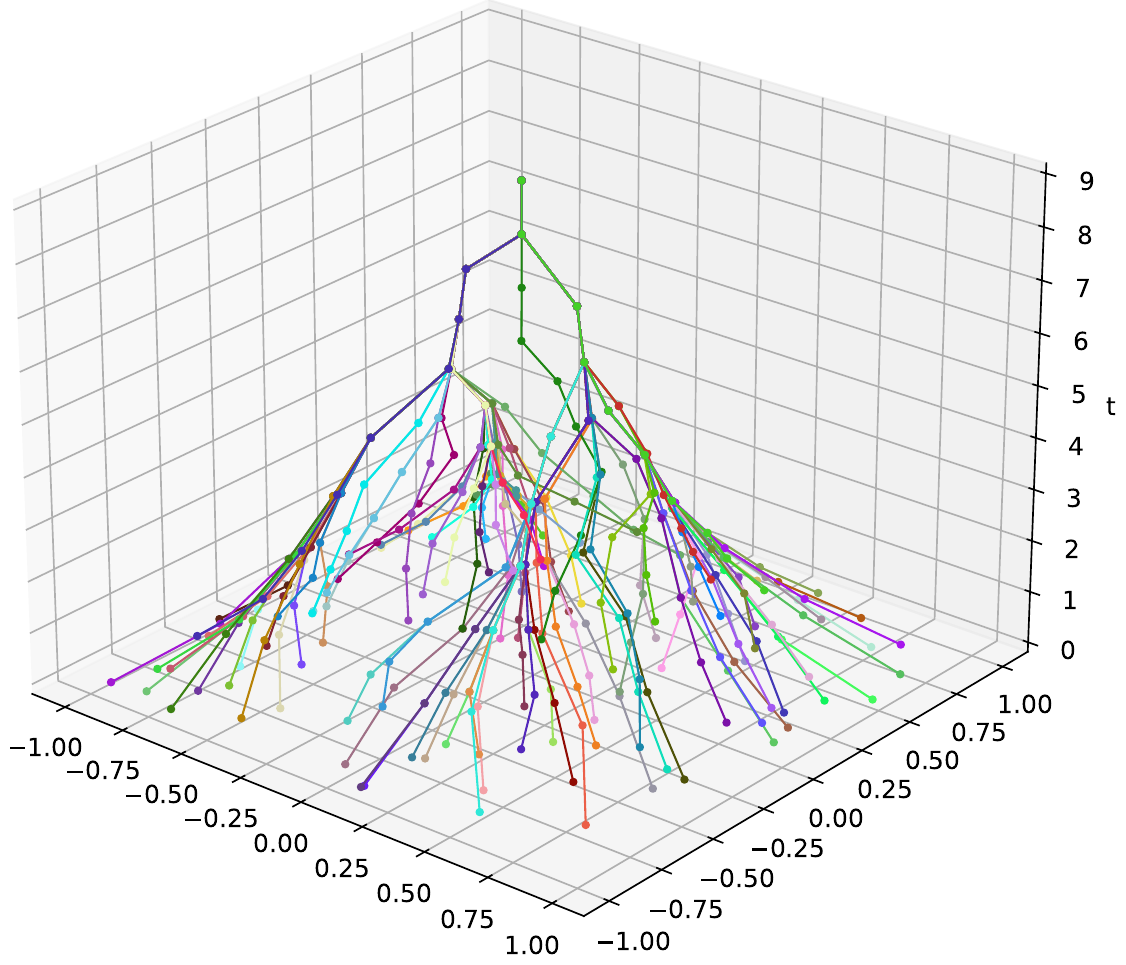}
         \caption{$p=0.1$}
     \end{subfigure}\hfill
     \begin{subfigure}[b]{0.33\textwidth}
         \centering
         \includegraphics[width=\textwidth]{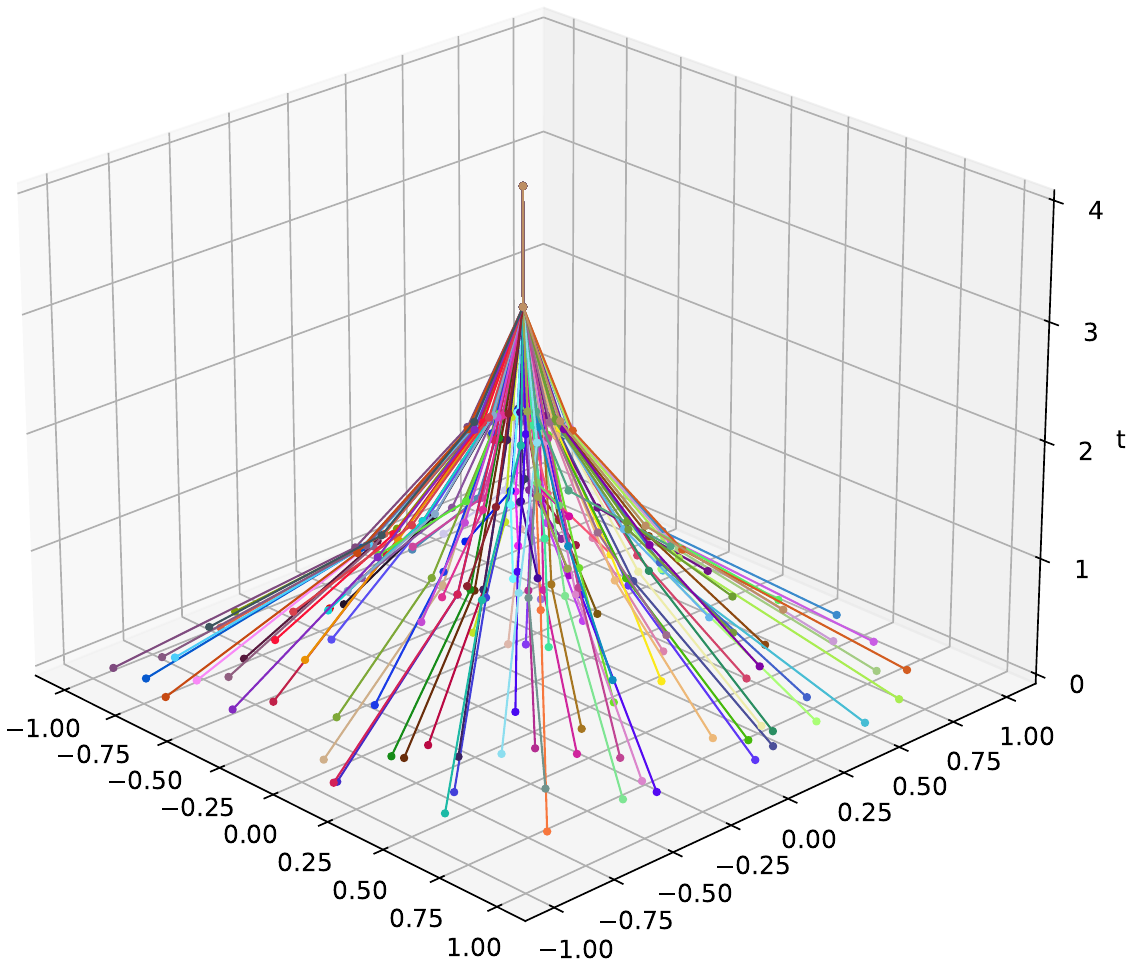}
         \caption{$p=0.5$}
    \end{subfigure}
     \caption{ODGP with different $p$}
     \label{fig.p-diff}
\end{figure*}

We conducted three tests for the ODGP with $n=100$ agents, using the public opinion~\eqref{eq.mean-value} and a confidence radius of $\varepsilon=0.5$. In these tests, we set $p=0.05$, $p=0.1$, and $p=0.2$, respectively, with identical initial conditions.
The actual consensus time, indeed, tends to decrease as $p$ is increasing (Fig.~\ref{fig.p-diff}),
although the dependence on $p$ needs further analysis.


\section{Conclusions}\label{sec.concl}

While the existence of group pressure driving individuals to conform with the majority has been widely accepted in social psychology since Asch's experiments, dynamic models describing this phenomenon remain scarce and are not yet well understood.
In this work, we generalize and analyze the model of consensus driven by social pressure, originally proposed in~\cite{cheng2019opinion}.
We demonstrate that consensus can still be reached even if conformity levels asymptotically decay, provided their decay is sufficiently slow. For uniformly positive conformity levels, we provide an explicit estimate of the consensus time and show that, unlike the estimate in~\cite{cheng2019opinion}, it is independent of group size. 

Several avenues remain open for future research. One such avenue is obtaining tighter estimates for the consensus time, which remains a non-trivial problem even for homogeneous agents with constant conformity levels $p$. An even more challenging direction is to examine scenarios where condition~\eqref{eq.series} is violated, such as when some agents are non-conformists with $p_i=0$. Some simulations provided in~\cite{cheng2019opinion} for the Hegselmann-Krause underlying dynamics illustrate that opinions converge to finite limits, though consensus is not guaranteed; however, formal mathematical proofs seem to be elusive. Another challenging problem is to analyze the dynamics of a system where multiple groups of agents are influenced by different public opinions (e.g., individuals receiving information from different media sources).

\section{Acknowledgements} 

The authors are grateful to the anonymous reviewers of the American Control Conference for their valuable suggestions on improving the text. We also thank Professor Serge Galam for drawing our attention to the majority rule models.

\bibliographystyle{IEEEtran}
\bibliography{bib_from_survey,bibnew}

\end{document}